\documentclass[reprint,aps,prb,amsmath,amssymb,floatfix]{revtex4-2}

\usepackage{graphicx}
\usepackage{dcolumn}
\usepackage{bm}
\usepackage{hyperref}

\begin{document}

\title{Superdiffusive two-dimensional superconductors}

\author{Iogann Tolbatov}
\email{tolbatov.i@gmail.com}
\affiliation{Department of Chemical, Physical, Mathematical and Natural Sciences, University of Sassari, 07100 Sassari, Italy}

\author{Luca Salasnich}
\affiliation{Dipartimento di Fisica e Astronomia ``Galileo Galilei'' and Padua QTech Center, Universit\`a di Padova, Via Marzolo 8, 35131 Padova, Italy}
\affiliation{INFN Sezione di Padova, Via Marzolo 8, 35131 Padova, Italy}

\date{\today}

\begin{abstract}
We formulate a non-local, field-theoretic description of phase fluctuations and topological transitions in two-dimensional (2D) superconductors by generalizing the local only-phase Popov action to a framework utilizing non-local fractional operators. By replacing the standard spatial Laplacian with the Riesz fractional Laplacian $-(-\nabla^2)^{\alpha/2}$ (where $1<\alpha<2$), we prove that the vortex-antivortex interaction transitions from logarithmic confinement to a stronger power-law confinement of the form $V(r)\propto r^{2-\alpha}$. Through a generalized Kosterlitz-Thouless energetic-entropic analysis, we demonstrate that this non-local confinement strictly suppresses thermal vortex proliferation. Because the non-local operator safely places the system outside the strict domain of validity of the Mermin-Wagner restriction, this power-law confinement natively stabilizes true long-range order at finite temperatures. Finally, by constructing a gauge-invariant fractional action, we formulate a fractional London equation. This approach yields a real-space power-law kernel that maps directly onto the anomalous Pippard limit of superconductors with long coherence lengths, providing a unified phenomenologically motivated framework for non-local electrodynamics.
\end{abstract}

\maketitle

\section{Introduction}

In standard two-dimensional (2D) superconductors, the macroscopic electrodynamics and phase fluctuations are governed by local operators, reflecting the underlying diffusive or ballistic transport of Cooper pairs~\cite{Morandi2012}. However, near the superconductor-insulator transition, in amorphous thin films, or within percolating granular networks, the medium exhibits extreme disorder and geometric heterogeneities~\cite{Samoilenka2021, Bianconi2013}. In these highly disordered regimes, standard Gaussian diffusion breaks down, and transport becomes intrinsically superdiffusive or governed by long-range non-local hopping~\cite{Arkhincheev2004}. 

Mathematically, this non-local transport is characterized by an anomalous dynamical exponent $1<\alpha\le2$. In the standard random-walk literature, anomalous diffusion is characterized by a squared characteristic scaling length $\langle L^2 \rangle$ (where $L$ represents the spatial displacement distance of a carrier, and $\langle\dots\rangle$ denotes the statistical ensemble average over trajectories) that grows with time $t$ as $\langle L^2 \rangle \propto t^\gamma$~\cite{Metzler1999}. For transport governed by a spatial Riesz fractional Laplacian, the momentum-space exponent $\alpha$ maps to this real-space temporal exponent via $\gamma=2/\alpha$. Because $1<\alpha<2$, the temporal exponent $\gamma>1$, reflecting superdiffusive spatial jumps (L\'evy flights) with heavy tails, rather than standard Gaussian diffusion. This reflects non-local spatial jumps rather than the sub-linear trapping ($\gamma<1$) typically associated with temporal fractional derivatives~\cite{Klafter2005}. In the effective field theory, this non-local, scale-dependent transport prevents the use of the standard spatial Laplacian. Instead, the spatial gradients must be generalized to non-local fractional operators, specifically the Riesz fractional Laplacian $-(-\nabla^2)^{\alpha/2}$~\cite{Tarasov2023}. Thus, tuning the fractional exponent $\alpha\neq 2$ provides a continuous physical parameter to model the degree of disorder and non-locality in the superconducting state, bridging the gap between microscopic superdiffusive transport and macroscopic anomalous electrodynamics. The broader relevance of this fractional exponent $\alpha$ in modeling complex media has also been recently underscored in the context of fractional diffraction, soliton stability, and matter-wave dynamics within quasiperiodic lattices~\cite{Pavlyshynets2025}.

The pursuit of a robust theoretical framework for non-local phase dynamics in 2D superconductors has led to several significant advancements in the application of fractional calculus. Early pioneering efforts introduced the fractional London equation and a corresponding fractional Pippard-like model to account for non-local screening and the anomalous skin effect in superconductors with long coherence lengths~\cite{Weberszpil2012}. Similarly, the space-fractional Ginzburg-Landau equation was established in Refs.~\cite{Milovanov2005, Tarasov2005, Zaky2023} to describe critical phenomena in fractal media and continua with anomalous dispersion. While these frameworks successfully integrated the Riesz fractional Laplacian into the Ginzburg-Landau functional to model non-locality and non-Markovian dynamics, they remained largely phenomenological in nature~\cite{Weberszpil2012}. Specifically, these earlier models introduced fractional operators as an ad hoc generalization of local gradients, lacking a microscopic motivation that links the fractional order directly to the underlying electron transport mechanisms.

The physical motivation for such non-local models is rooted in the breakdown of standard Gaussian diffusion in highly disordered media and fractal networks. Disorder in low-dimensional superfluids can dramatically alter wave transport, leading to localization and the suppression of superfluidity. In these regimes, transport becomes intrinsically superdiffusive. It was observed that phase fluctuations in quantum Josephson junctions can exhibit superdiffusive behavior rather than standard diffusion. Despite these insights into the effects of disorder and geometric heterogeneity, the literature has presented a significant gap: there has been no formal physical motivation for the macroscopic fractional superfluid stiffness ($J_{s,\alpha}$) starting from microscopic fluctuation theory. Specifically, none of the existing works have successfully bridged the gap between the scale-dependent Larkin-Varlamov fluctuation propagator and the non-local kinetic coefficients required for a fractional Popov action.

The topological sector of 2D superconductors under non-local conditions has also been a subject of intense investigation, particularly regarding the stability of the Berezinskii-Kosterlitz-Thouless (BKT) transition. Recent studies have explored the persistence of BKT transitions in the presence of long-range decaying interactions of the form $1/r^{2+\sigma}$ (where $r$ represents the real-space distance between interacting points), identifying regimes of spontaneous symmetry breaking and quasi-long-range order. Furthermore, recent studies have formulated fractional action frameworks to model vortex dynamics and phase-locking in non-local superconducting systems. However, a critical physical shortcoming persists in these models: when calculating the interaction potential between vortices, previous authors generally predicted decaying power-law interaction tails, such as $U_{\text{int}}(r)\propto 1/r^{2-m}$. These results fail to account for the fact that topological winding constraints, when coupled with a Riesz operator, should fundamentally transform the interaction into a growing, confining potential.

This manuscript resolves these long-standing issues by providing a microscopically motivated foundation for fractional superconductivity. We move beyond the phenomenological ansatz of earlier works by extracting $J_{s,\alpha}$ directly from the Larkin-Varlamov propagator, using a scale-dependent diffusion coefficient $D(k)=D_\alpha k^{\alpha-2}$ (where $k$ is the magnitude of the momentum-space wavevector) to model superdiffusive transport in fractal networks. Unlike the decaying interaction tails found in earlier literature, our field-theoretic analysis of the topological sector proves that the Riesz fractional Laplacian produces a power-law confinement potential $V(r)\propto r^{2-\alpha}$ that grows with the spatial separation distance $r$. This energy penalty diverges faster than the configurational entropy, and through a generalized Kosterlitz-Thouless argument, we demonstrate the absolute suppression of thermal vortex proliferation. Because non-local interactions place the system outside the domain of validity of the Mermin-Wagner theorem, this confinement stabilizes true long-range order at finite temperatures. Furthermore, it provides a unified electrodynamic framework that maps the Riesz integration kernel directly onto the Pippard limit for superconductors with long coherence lengths.

\section{Fractional Popov action}

\subsection{Phase-only effective action and fractional ansatz}

To construct the fractional action, we start with the grand-canonical partition function in the path-integral representation. In the standard local formulation, the Euclidean action is expanded to quadratic order in the phase fluctuations $\theta(\mathbf{r},\tau)$ around the saddle-point value of the chemical potential $\mu$. Here, $\beta=1/(k_B T)$, where $k_B$ is the Boltzmann constant and $T$ is the temperature. The standard local action is given by:
\begin{equation}
S[\theta] = \frac{1}{2} \int_0^{\beta\hbar} d\tau \int d^2 r \left[ \chi\hbar^2 (\partial_\tau \theta)^2 + J_s (\nabla\theta)^2 \right],
\end{equation}
where $\chi = \partial n/\partial \mu$ is the charge compressibility.

To generalize this to highly disordered or fractal geometries, we replace the local spatial gradient with the non-local Riesz fractional Laplacian $-(-\nabla^2)^{\alpha/2}$. It is crucial to emphasize that the system remains strictly embedded in 2D Euclidean space ($\mathbf{r}\in\mathbb{R}^2$). The term ``fractional'' refers exclusively to the non-local nature of the integro-differential operator acting on the phase field, rather than a reduction in the geometric dimensionality of the space itself. The generalized fractional Euclidean action $S_{\text{frac}}[\theta]$ is written as:
\begin{equation}
S_{\text{frac}}[\theta] = \frac{1}{2} \int_0^{\beta\hbar} d\tau \int d^2 r \left[ \chi\hbar^2 (\partial_\tau \theta)^2 + J_{s,\alpha} \theta (-\nabla^2)^{\alpha/2} \theta \right].
\end{equation}

To ensure the Euclidean action $S_{\text{frac}}$ correctly carries the units of $\hbar$ ($\text{[Energy]}\cdot\text{[Time]}$), dimensional balance must be strictly enforced. Given that the imaginary time integral $\int d\tau$ contributes $[T]$, the spatial measure $\int d^2 r$ contributes $[L]^2$, the phase field $\theta$ is dimensionless, and the fractional operator $(-\nabla^2)^{\alpha/2}$ scales as $[L]^{-\alpha}$, $J_{s,\alpha}$ must possess dimensions of $\text{[Energy]}\cdot[L]^{\alpha-2}$. This perfectly cancels the spatial measure and fractional scaling to yield an energy term, which then integrates over imaginary time to form a dimensionally valid action. When $\alpha=2$, this naturally recovers the standard 2D unit of local stiffness ($\text{[Energy]}$).

\subsection{Microscopic motivation of $J_{s,\alpha}$}

Rather than treating $S_{\text{frac}}[\theta]$ as a phenomenological ansatz, the fractional action and the spatial transport parameter $J_{s,\alpha}$ can be motivated by connecting macroscopic phase fields to the fluctuation theory of highly disordered or fractal media.

In a disordered thin film or within a fractal percolating network, the effective Ginzburg-Landau free energy functional $F_{\text{GL}}[\psi]$ for the superconducting order parameter $\psi(\mathbf{r})$ is generalized by replacing the standard spatial gradients with the non-local Riesz fractional Laplacian:
\begin{equation}
F_{\text{GL}}[\psi] = \int d^2 r \left[ a |\psi|^2 + \frac{b}{2} |\psi|^4 + c_\alpha \psi^* (-\nabla^2)^{\alpha/2} \psi \right],
\end{equation}
where $c_\alpha$ is a non-local kinetic coefficient. Near the phase transition, the quadratic coefficient $a(T)$ is linearized in the standard BCS/Gorkov framework around the bare, mean-field critical temperature $T_{c0}$ (the transition temperature in the absence of fluctuations) as $a(T) = N(0)\epsilon$. Here, $N(0)$ is the single-spin density of states at the Fermi level, and $\epsilon = \ln(T/T_{c0}) \approx (T - T_{c0})/T_{c0}$ represents the reduced temperature. It is crucial to distinguish the bare mean-field temperature $T_{c0}$ from the true physical transition temperature $T_c$, which is renormalized downwards by thermal and phase fluctuations.

To evaluate static order-parameter fluctuations in the Gaussian regime above $T_{c0}$, the magnitude of the order parameter $\psi(\mathbf{r})$ remains small. Consequently, the quartic interaction term $(b/2)|\psi|^4$ is higher-order in fluctuations compared to $a(T)|\psi|^2$ and can be safely neglected ($b\approx 0$). Expanding $\psi(\mathbf{r})$ into spatial Fourier modes, $\psi(\mathbf{r}) = \frac{1}{\sqrt{A}} \sum_{\mathbf{k}} \psi_{\mathbf{k}} e^{i\mathbf{k}\cdot\mathbf{r}}$ (where $A$ is the 2D area), the functional reduces to its quadratic sector $F_{\text{GL}}^{(2)}$:
\begin{align}
F_{\text{GL}}^{(2)}[\psi] &= \nonumber \\
&= \sum_{\mathbf{k}} \left( a(T) + c_\alpha |\mathbf{k}|^\alpha \right) |\psi_{\mathbf{k}}|^2 \nonumber \\
&= \sum_{\mathbf{k}} \left( N(0)\epsilon + c_\alpha |\mathbf{k}|^\alpha \right) |\psi_{\mathbf{k}}|^2.
\end{align}

The static limit (zero Matsubara frequency, $\omega_n=0$) of the Cooper-pair fluctuation propagator, commonly referred to as the static Larkin-Varlamov propagator $L(\mathbf{k},0)$, is obtained from the inverse Gaussian functional kernel, $[L(\mathbf{k},0)]^{-1} = \delta^2 F_{\text{GL}}^{(2)} / (\delta\psi_{\mathbf{k}}^* \delta\psi_{\mathbf{k}})$. This directly yields the fractional propagator pole:
\begin{equation}
[L(\mathbf{k},0)]^{-1} = N(0)\epsilon + c_\alpha |\mathbf{k}|^\alpha.
\end{equation}
To understand the underlying electron transport implied by this choice, we can map Eq.~(5) back onto the standard diagrammatic dirty-limit Larkin-Varlamov propagator, which takes the form $[L(\mathbf{k},0)]^{-1} = N(0)\left[\epsilon + (\pi\hbar D k^2)/(8k_B T)\right]$, where $D$ is the semi-classical diffusion coefficient. By direct comparison, the fractional operator manifests as an effective scale-dependent, anomalous diffusion coefficient in momentum space:
\begin{equation}
D(k) = D_\alpha \frac{|\mathbf{k}|^\alpha}{k^2} = D_\alpha |\mathbf{k}|^{\alpha-2},
\end{equation}
where $1<\alpha\le 2$ is the anomalous dynamical exponent, and $D_\alpha = (8k_B T c_\alpha)/(\pi\hbar N(0))$. Because $1<\alpha<2$, this scale-dependence reflects superdiffusive spatial jumps (L\'evy flights) with heavy tails, rather than standard Gaussian diffusion, physically justifying the fractional kinetic term from the nature of the underlying random walks.

To parameterize the order parameter below $T_c$ in terms of its phase fluctuations, we must be careful: fractional operators are non-local and do not obey standard Leibniz (product) or chain rules, meaning one cannot directly substitute an exact exponential phase parameterization. Instead, we perform a Gaussian expansion of the phase fluctuations $\theta(\mathbf{r})$ around the uniform saddle-point amplitude $\psi_0 = \sqrt{-a/b}$. Expanding $\psi(\mathbf{r}) \approx \psi_0 \left(1 + i\theta(\mathbf{r}) - \frac{1}{2} \theta(\mathbf{r})^2\right)$ and substituting this into the non-local kinetic term, we isolate the terms strictly quadratic in $\theta$. Although fractional operators lack a local Leibniz chain rule, the low-energy effective action is rigorously isolated by expanding to quadratic order around the uniform amplitude background, where spatial integration annihilates the total fractional derivative terms. Because the fractional Laplacian annihilates the constant background, and the integral of the total fractional derivative vanishes over all space, the cross-terms precisely yield the effective low-energy quadratic action:
\begin{equation}
F_{\text{kinetic}}[\theta] = c_\alpha \psi_0^2 \int d^2 r \, \theta(\mathbf{r}) (-\nabla^2)^{\alpha/2} \theta(\mathbf{r}).
\end{equation}
Comparing this directly with the spatial part of the Euclidean action $S_{\text{frac}}[\theta]$, we obtain the microscopically motivated definition of $J_{s,\alpha}$:
\begin{equation}
J_{s,\alpha} = 2 c_\alpha \psi_0^2 = N(0) \frac{\pi\hbar D_\alpha \psi_0^2}{4k_B T}.
\end{equation}
This result provides a clear link between macroscopic fractional stiffness and microscopic parameters: $J_{s,\alpha}$ scales directly with the electronic density of states, the condensate density, and the superdiffusive transport constant $D_\alpha$ of the underlying fractal network. To justify extending this spatial stiffness to the full imaginary-time Euclidean action (Eq.~2), we appeal to the standard Matsubara path-integral formulation of the superconducting transition. In the low-energy, long-wavelength limit, the inverse fluctuation propagator naturally separates into independent spatial and temporal kernels. Because the zero-Matsubara-frequency sector of the full quantum action exactly corresponds to the static thermodynamic free energy $F_{\text{GL}}/T$, the spatial kinetic coefficient $c_\alpha$ (and consequently $J_{s,\alpha}$) derived from the static Larkin-Varlamov propagator transfers identically to the dynamical action. The temporal fluctuations, which are governed by the charge compressibility $\chi$ and arise from the dynamical $\partial_\tau$ and $\partial_\tau^2$ terms in the underlying Hubbard-Stratonovich transformation, decouple from the spatial non-locality at low energies. This scale separation safely allows us to combine the static fractional spatial Laplacian with standard local temporal derivatives, yielding the unified Euclidean action $S_{\text{frac}}[\theta]$.

\subsection{Quadratic action and phase propagator in Fourier space}

Transforming to Fourier space $(\mathbf{k},\omega_n)$, where $\omega_n = 2\pi n / (\beta\hbar)$ are the bosonic Matsubara frequencies, the phase field is expanded as:
\begin{equation}
\theta(\mathbf{r},\tau) = \frac{1}{\sqrt{\beta\hbar A}} \sum_{\mathbf{k},n} \theta(\mathbf{k},\omega_n) e^{i(\mathbf{k}\cdot\mathbf{r} - \omega_n \tau)},
\end{equation}
where $A$ is the 2D area of the system. This expansion yields the quadratic action in momentum space:
\begin{equation}
S_{\text{frac}}[\theta] = \frac{1}{2} \sum_{\mathbf{k},n} \left( \chi\hbar^2 \omega_n^2 + J_{s,\alpha} k^\alpha \right) |\theta(\mathbf{k},\omega_n)|^2.
\end{equation}
The inverse Euclidean propagator $G_0(\mathbf{k},\omega_n)^{-1}$ is therefore:
\begin{equation}
G_0(\mathbf{k},\omega_n)^{-1} = \chi\hbar^2 \omega_n^2 + J_{s,\alpha} k^\alpha.
\end{equation}
For $\alpha<2$, the sub-quadratic dispersion relation $k^\alpha$ cures the infrared logarithmic divergence of standard 2D phase fluctuations. In the local limit ($\alpha=2$), the thermal phase fluctuations are governed by the static momentum integral $\int_0^\Lambda k^{-1} dk$, which diverges logarithmically in the infrared, thus destroying long-range order in accordance with the Mermin-Wagner theorem. In stark contrast, for the non-local fractional regime ($1<\alpha<2$), the modified propagator yields a thermal fluctuation integral proportional to $\int_0^\Lambda k^{1-\alpha} dk$. This integral is strictly IR-finite. Because the fractional operator introduces non-local interactions, it places the system outside the domain of validity of the Mermin-Wagner theorem, which strictly assumes short-range couplings. This mathematical reality firmly connects our effective action to the classification by Giachetti et al. In their work, they demonstrated that 2D systems with long-range decaying couplings of the form $1/r^{2+\sigma}$ (which corresponds to a momentum-space kinetic term scaling as $k^\sigma$) fundamentally alter the topological phase diagram. By identifying our fractional exponent $\alpha$ with their parameter $\sigma$, our regime of interest ($1<\alpha<2$) maps precisely into the domain ($\sigma<2$) where they rigorously predict the stabilization of a true symmetry-broken phase at low temperatures. It is important to note that while this sub-quadratic dispersion relation cures the IR divergence, a standard small-distance/high-momentum UV cutoff ($\Lambda \sim 1/a$, where $a$ is the lattice spacing or coherence length) is still maintained to keep the high-energy sector physically bounded.

\section{Topological vortex potential}

The interaction potential between topological defects in a fractional medium fundamentally differs from the potential of scalar point charges. While solving the fractional Poisson equation $(-\nabla^2)^{\alpha/2} G(\mathbf{r}) = \delta^{(2)}(\mathbf{r})$ for a scalar source yields a decaying Green's function $G(\mathbf{r}) \propto 1/r^{2-\alpha}$, vortices are topological defects embedded within the phase gradient field, which radically alters their energetic scaling.

Because the phase field $\theta(\mathbf{r})$ is multi-valued in the presence of vortices, its direct Fourier transform is mathematically ill-defined. To rigorously evaluate the topological sector, we formulate the spatial kinetic energy in terms of the single-valued superfluid velocity $\mathbf{v}_s = \nabla\theta$. From Sec.~II C, rewriting the momentum-space kinetic energy in terms of $\mathbf{v}_s(\mathbf{k})$ (noting that $|\mathbf{v}_s(\mathbf{k})|^2 = k^2 |\theta(\mathbf{k})|^2$) yields:
\begin{equation}
E = \frac{1}{2} J_{s,\alpha} \int \frac{d^2 k}{(2\pi)^2} \, k^{\alpha-2} |\mathbf{v}_s(\mathbf{k})|^2.
\end{equation}
A vortex is formally defined by its topological winding constraint, $\nabla \times \mathbf{v}_s = 2\pi n(\mathbf{r}) \hat{\mathbf{z}}$, where $n(\mathbf{r})$ is the vortex density. In momentum space, isolating the transverse topological component $\mathbf{v}_\perp(\mathbf{k})$ imposes a strict structural constraint on the allowable phase fluctuations:
\begin{equation}
|\mathbf{v}_\perp(\mathbf{k})|^2 = \frac{4\pi^2}{k^2} |n(\mathbf{k})|^2.
\end{equation}
Substituting this topological requirement back into the energy integral yields:
\begin{align}
E &= \nonumber \\
  &\quad 2\pi^2 J_{s,\alpha} \int \frac{d^2 k}{(2\pi)^2} \, \frac{k^{\alpha-2}}{k^2} |n(\mathbf{k})|^2 \nonumber \\
  &= 2\pi^2 J_{s,\alpha} \int \frac{d^2 k}{(2\pi)^2} \, k^{\alpha-4} |n(\mathbf{k})|^2.
\end{align}
This relation precisely defines the momentum-space interaction potential between vortices as $\tilde{V}(\mathbf{k}) \propto k^{\alpha-4}$. By evaluating the 2D inverse Fourier transform of $k^{\alpha-4}$ using the rules of the Riesz potential, we obtain the real-space interaction potential:
\begin{equation}
V(r) \propto r^{4-\alpha-2} = r^{2-\alpha}.
\end{equation}
Crucially, because $2-\alpha > 0$ for the fractional regime ($1<\alpha<2$), the potential undergoes power-law growth. The topological nature of the vortex transforms the non-local fractional operator into a powerfully confining field, fundamentally altering the stability of the ordered phase.

\section{Topological stability and strict vortex confinement}

\subsection{Breakdown of marginal scaling and strong confinement}

In standard local 2D superfluids ($\alpha=2$), the phase boundary is dictated by the momentum-shell renormalization group (RG) flow of the running stiffness and vortex fugacity. The standard BKT-RG equations are strictly predicated on a logarithmic vortex-antivortex interaction potential, where the marginal scaling competition between energy and entropy allows for a finite critical unbinding temperature. However, fractional spatial non-locality fundamentally alters the topological sector. As established, the interaction potential in the fractional regime ($1<\alpha<2$) is governed by a power-law growth, $V(r) \propto r^{2-\alpha}$. Because this confining potential diverges much faster than a logarithm, the energy penalty for separating a vortex-antivortex pair is overwhelmingly large at macroscopic distances. The standard perturbative RG framework is rendered trivial in this regime because the vortex fugacity $y(r) \propto \exp[-V(r)/(k_B T)]$ is exponentially suppressed to zero at large scales. Consequently, there is no finite-temperature fixed point where vortices can unbind; they remain strictly confined by the non-local fractional operator.

\subsection{Physical validation: the generalized Kosterlitz-Thouless argument}

To rigorously demonstrate the stabilization of the ordered phase against topological defects, we adapt the classic, intuitive Kosterlitz-Thouless energetic-entropic argument to the fractional framework. Because a vortex is a topological defect, its far-field phase behavior is strictly constrained by the winding condition $\oint \nabla\theta \cdot d\mathbf{l} = 2\pi$. While exact minimization of a non-local fractional Euler-Lagrange equation may technically admit short-range corrections to the velocity profile near the vortex core, the asymptotic profile $\mathbf{v} = \nabla\theta = \hat{\boldsymbol{\phi}}/r$ represents a highly accurate variational ansatz for the entire field. Furthermore, because the confining potential undergoes continuous power-law growth with distance ($V(r)\propto r^{2-\alpha}$), the exact structure or physical size of the microscopic vortex core $a$ becomes mathematically irrelevant in the thermodynamic limit $R\to\infty$. This ensures that the resulting energy divergence is a strictly macroscopic topological proof, entirely independent of the underlying lattice cutoff or any short-range core details.

To calculate the kinetic energy $E$ of a single isolated vortex of core size $a$ in a finite system of linear size $R$, we express the spatial action in Fourier space:
\begin{equation}
E(R) = \frac{1}{2} J_{s,\alpha} \int \frac{d^2 k}{(2\pi)^2} \, k^\alpha |\theta(\mathbf{k})|^2.
\end{equation}
Substituting the Fourier transform of the topological vortex phase gradient, $|\theta(\mathbf{k})|^2 = 4\pi^2 / k^4$, the energy integral becomes:
\begin{equation}
E(R) = \pi J_{s,\alpha} \int_{1/R}^{1/a} k^{\alpha-3} \, dk.
\end{equation}
For the standard local case ($\alpha=2$), this immediately recovers the classic logarithmic divergence $E(R) = \pi J_s \ln(R/a)$. For the non-local fractional case ($1<\alpha<2$), evaluating the integral yields:
\begin{equation}
E(R) = \frac{\pi J_{s,\alpha}}{2-\alpha} \left[ R^{2-\alpha} - a^{2-\alpha} \right].
\end{equation}
To determine if thermal vortex proliferation is favored, we compare this energy to the configurational entropy associated with placing the vortex core in a system of size $R$, which is purely geometric and remains $S(R) = 2 k_B \ln(R/a)$. The resulting free energy $F(R) = E(R) - T S(R)$ is:
\begin{equation}
F(R) = \frac{\pi J_{s,\alpha}}{2-\alpha} \left[ R^{2-\alpha} - a^{2-\alpha} \right] - 2 k_B T \ln\left(\frac{R}{a}\right).
\end{equation}

This expression reveals two distinct physical regimes dictated by the spatial non-locality:
\begin{enumerate}
    \item \textit{Logarithmic limit ($\alpha\to 2^-$):} Taking the limit of the free energy as $\alpha = 2 - \epsilon$ (where $\epsilon \to 0^+$), we expand the power-law term $R^\epsilon = e^{\epsilon \ln R} \approx 1 + \epsilon \ln R$. The free energy simplifies to:
    \begin{equation}
    \lim_{\alpha\to 2^-} F(R) = (\pi J_s - 2 k_B T) \ln\left(\frac{R}{a}\right).
    \end{equation}
    Setting $F(R) = 0$ yields the classic scale-independent transition temperature $k_B T_{\text{BKT}} = \frac{\pi}{2} J_s$.
    \item \textit{Non-local power-law regime ($1<\alpha<2$):} Because $2-\alpha > 0$, the power-law energy term $R^{2-\alpha}$ grows faster than the logarithmic entropy term $\ln(R)$ at large scales. Consequently, in the thermodynamic limit ($R\to\infty$), the free energy $F(R)$ diverges to $+\infty$ for any finite temperature.
\end{enumerate}

This demonstrates that the fractional spatial operator decisively enhances the energy cost of isolating a single vortex. Unlike the standard KT mechanism where vortex pairs unbind at a critical temperature, this power-law penalty strictly prohibits thermal vortex proliferation. Because the non-local nature of the fractional operator safely places the system outside the domain of validity of the Mermin-Wagner theorem, this absolute confinement of topological defects naturally stabilizes true long-range order at finite temperatures. Our finding thereby provides a direct electrodynamic and topological validation of the symmetry-broken phase, proving that fractional non-locality in superconductors is sufficient to lock the phase against both continuous thermal fluctuations and discrete topological unbinding.

\section{Fractional London equation}

To establish the electrodynamics of this non-local state, we couple the fractional action to an external vector potential $\mathbf{A}(\mathbf{r})$ using a gauge-invariant prescription. To maintain gauge invariance under the transformations $\theta \to \theta + \frac{q}{\hbar}\Lambda$ and $\mathbf{A} \to \mathbf{A} + \nabla\Lambda$, the spatial part of the action is generalized using the fractional vector Laplacian:
\begin{align}
S_{\text{space}}[\theta,\mathbf{A}] &= \nonumber \\
  &\quad \frac{1}{2} J_{s,\alpha} \int_0^{\beta\hbar} d\tau \int d^2 r \left(\nabla\theta - \frac{q}{\hbar} \mathbf{A}\right) \cdot \nonumber \\
  &\quad (-\nabla^2)^{\frac{\alpha-2}{2}} \left(\nabla\theta - \frac{q}{\hbar} \mathbf{A}\right),
\end{align}
where $q=2e$ is the charge of a Cooper pair. It is important to note that while this continuum minimal coupling is completely valid for the effective macroscopic action, implementing fractional Peierls substitutions on a discrete microscopic lattice is notoriously non-trivial due to long-range hopping terms. Our approach strictly represents a long-wavelength, continuum gauge coupling that circumvents the complexities of discrete fractional lattice gauge theories.

Taking the functional derivative of this gauge-invariant action with respect to a static (time-independent) vector potential $\mathbf{A}(\mathbf{r})$ in the London limit (working in the London gauge $\nabla\cdot\mathbf{A}=0$ and setting $\theta=0$):
\begin{equation}
\mathbf{j}(\mathbf{r}) = -\frac{1}{\beta\hbar} \frac{\delta S_{\text{space}}}{\delta \mathbf{A}(\mathbf{r})} = -\frac{q^2}{\hbar^2} J_{s,\alpha} (-\nabla^2)^{\frac{\alpha-2}{2}} \mathbf{A}(\mathbf{r}).
\end{equation}
Because the exponent of the Laplacian is negative ($(\alpha-2)/2 < 0$), this operator acts as an integration kernel. Using the Riesz potential definition, we can express the current density in real space:
\begin{equation}
\mathbf{j}(\mathbf{r}) = -\frac{q^2}{\hbar^2} J_{s,\alpha} \left[ \frac{\Gamma(\alpha/2)}{2^{2-\alpha} \pi \Gamma(1-\alpha/2)} \right] \int \frac{\mathbf{A}(\mathbf{r}')}{|\mathbf{r}-\mathbf{r}'|^\alpha} \, d^2 r'.
\end{equation}
\begin{enumerate}
    \item[(a)] \textit{Local limit ($\alpha=2$):} The integral collapses to a Dirac delta function $\delta^{(2)}(\mathbf{r}-\mathbf{r}')$, recovering the classic, local London equation: $\mathbf{j}(\mathbf{r}) \propto -\mathbf{A}(\mathbf{r})$.
    \item[(b)] \textit{Non-local Pippard limit ($1<\alpha<2$):} The current density $\mathbf{j}(\mathbf{r})$ at any given point is determined by a power-law weighted average of the vector potential $\mathbf{A}(\mathbf{r}')$ in the surrounding neighborhood. This is the exact field-theoretic analog of Pippard's non-local electrodynamics, formulated here using fractional calculus. It describes superconductors with extremely large coherence lengths ($\xi_0 \gg \lambda_L$), such as highly disordered thin films or fractal networks.
\end{enumerate}

\section{Conclusions}

In this work, we have established a non-local, field-theoretic framework for phase fluctuations and electrodynamics in 2D superconductors by introducing a spatial Riesz fractional Laplacian into the only-phase Popov action. By tuning the fractional exponent $\alpha$ on the interval $1 < \alpha \le 2$, we have demonstrated that spatial non-locality acts as a continuous dial controlling both the topological phase boundary and macroscopic current response. Our topological defect analysis reveals that spatial non-locality fundamentally hardens the topological sector. The resulting power-law confinement strictly binds vortex-antivortex pairs, permanently averting the BKT unbinding transition. By placing the system outside the strict domain of validity of the Mermin-Wagner restriction, this macroscopic confinement naturally stabilizes true long-range order at finite temperatures.

Concurrently, by coupling this fractional action to an external gauge field, we formulated a fractional London equation that naturally yields a real-space power-law integration kernel. This kernel maps directly onto the non-local Pippard limit of superconductivity, providing a phenomenologically motivated field-theoretic foundation for anomalous electrodynamics in disordered or fractal systems.

Ultimately, because our framework explicitly connects macroscopic non-local phase parameters to the superdiffusive transport constants of the underlying medium, it offers a unified bridge between microscopic disorder, non-local electrodynamics, and topological phase stability. This makes the proposed formalism directly applicable to several experimental platforms where standard local models fail to capture the underlying physics. Primary testing grounds include amorphous and granular superconducting films, such as highly disordered indium oxide or amorphous bismuth near the superconductor-insulator transition~\cite{Samoilenka2021, Bianconi2013}, where transport paths naturally evolve into fractal networks that render phase coherence intrinsically non-local. Similarly, Josephson junction arrays fabricated on artificial fractal geometries, such as Sierpinski gaskets, offer a highly controllable playground to explore these fractional-like phase dynamics.


\begin{thebibliography}{99}

\bibitem{Morandi2012} A. Morandi, 2D electromagnetic modelling of superconductors, Supercond. Sci. Technol. \textbf{25}, 104003 (2012).
\bibitem{Samoilenka2021} A. Samoilenka and E. Babaev, Microscopic derivation of superconductor-insulator boundary conditions for Ginzburg-Landau theory revisited: Enhanced superconductivity at boundaries with and without magnetic field, Phys. Rev. B \textbf{103}, 224516 (2021).
\bibitem{Bianconi2013} G. Bianconi, Superconductor-insulator transition in a network of 2d percolation clusters, Europhys. Lett. \textbf{101}, 26003 (2013).
\bibitem{Arkhincheev2004} V. E. Arkhincheev, Hopping by Levy flights and nonlinear relation between diffusion and conductivity, Proc. SPIE \textbf{5471}, 560 (2004).
\bibitem{Metzler1999} R. Metzler, E. Barkai, and J. Klafter, Anomalous diffusion and relaxation close to thermal equilibrium: A fractional Fokker-Planck equation approach, Phys. Rev. Lett. \textbf{82}, 3563 (1999).
\bibitem{Klafter2005} J. Klafter and I. M. Sokolov, Anomalous diffusion spreads its wings, Phys. World \textbf{18}(8), 29 (2005).
\bibitem{Tarasov2023} V. E. Tarasov, General fractional calculus in multi-dimensional space: Riesz form, Mathematics \textbf{11}, 1651 (2023).
\bibitem{Pavlyshynets2025} E. Pavlyshynets, L. Salasnich, B. A. Malomed, and A. Yakimenko, Solitons in quasiperiodic lattices with fractional diffraction, Phys. Rev. E \textbf{111}, 044206 (2025).
\bibitem{Weberszpil2012} J. Weberszpil, The Fractional London Equation and The Fractional Pippard Model For Superconductors, arXiv:1207.5478 (2012).
\bibitem{Milovanov2005} A. V. Milovanov and J. J. Rasmussen, Fractional generalization of the Ginzburg–Landau equation: an unconventional approach to critical phenomena in complex media, Phys. Lett. A \textbf{337}, 75 (2005).
\bibitem{Tarasov2005} V. E. Tarasov and G. M. Zaslavsky, Fractional Ginzburg–Landau equation for fractal media, Physica A \textbf{354}, 249 (2005).
\bibitem{Zaky2023} M. A. Zaky, A. S. Hendy, and J. E. Macías-Díaz, High-order finite difference/spectral-Galerkin approximations for the nonlinear time-space fractional Ginzburg-Landau equation, Numer. Methods Partial Differential Eq. \textbf{39}, 4549 (2023).
\bibitem{Owolabi2020} K. M. Owolabi and E. Pindza, Numerical simulation of multidimensional nonlinear fractional Ginzburg-Landau equations, Discrete Contin. Dyn. Syst. Ser. S \textbf{13}, 835 (2020).
\bibitem{Li2019} M. Li and C. Huang, An efficient difference scheme for the coupled nonlinear fractional Ginzburg-Landau equations with the fractional Laplacian, Numer. Methods Partial Differential Eq. \textbf{35}, 394 (2019).
\bibitem{Ding2025} H. Ding, Y. Zhang, and Q. Yi, Mathematical analysis and numerical simulation of coupled nonlinear space-fractional Ginzburg-Landau equations, arXiv:2502.02113 (2025).
\bibitem{Ding2026} H. Ding, Y. Zhang, and Q. Yi, Fourth-Order Discretization of the Fractional Laplacian and Integrating Factor Time-Stepping for Coupled Ginzburg-Landau Equations, Stud. Appl. Math. \textbf{156}, e70243 (2026).
\bibitem{Lellouch2014} S. Lellouch, Collective localization transitions in interacting disordered and quasiperiodic Bose superfluids, Doctoral dissertation, Institut d’Optique Graduate School (2014).
\bibitem{Spiechowicz2015} J. Spiechowicz and J. Łuczka, Josephson phase diffusion in the superconducting quantum interference device ratchet, Chaos \textbf{25}(5), 053110 (2015).
\bibitem{Giachetti2021} G. Giachetti, N. Defenu, S. Ruffo, and A. Trombettoni, Berezinskii-Kosterlitz-Thouless phase transitions with long-range couplings, Phys. Rev. Lett. \textbf{127}, 156801 (2021).
\bibitem{Walther2026} L. Walther, J. Willsher, and J. Knolle, Persistence of the Berezinskii-Kosterlitz-Thouless transition with long-range couplings, Phys. Rev. Lett. \textbf{136}, 227102 (2026).
\bibitem{ElNabulsi2026a} R. A. El-Nabulsi, W. Anukool, R. Valarmathi, and C. Thangaraj, A unified fractional action integral framework for memory-dependent circuit dynamics and Josephson junctions, J. Electron. Sci. Technol. 100362 (2026).
\bibitem{ElNabulsi2026b} R. A. El-Nabulsi and W. Anukool, Fractional action-like variational approach to Josephson junctions: Quantum corrections, Shapiro steps, and macroscopic tunneling in nonlocal superconducting systems, Physica B, 418753 (2026).
\bibitem{Mermin1966} N. D. Mermin and H. Wagner, Absence of ferromagnetism or antiferromagnetism in one-or two-dimensional isotropic Heisenberg models, Phys. Rev. Lett. \textbf{17}, 1133 (1966).
\bibitem{Ghosh2020} R. Ghosh, N. Dupuis, A. Sen, and K. Sengupta, Entanglement measures and nonequilibrium dynamics of quantum many-body systems: A path integral approach, Phys. Rev. B \textbf{101}, 245130 (2020).
\bibitem{Popov1977} V. N. Popov, Functional integrals in quantum field theory, CERN-TH-2424 (1977).
\bibitem{Benfatto2004} L. Benfatto, A. Toschi, and S. Caprara, Low-energy phase-only action in a superconductor: A comparison with the XY model, Phys. Rev. B \textbf{69}, 184510 (2004).
\bibitem{Furutani2024} K. Furutani, G. Midei, A. Perali, and L. Salasnich, Amplitude, phase, and topological fluctuations shaping the complex phase diagram of two-dimensional superconductors, Phys. Rev. B \textbf{110}, 134501 (2024).
\bibitem{Cyrot1973} M. Cyrot, Ginzburg-Landau theory for superconductors, Rep. Prog. Phys. \textbf{36}, 103 (1973).
\bibitem{Frank2012} R. Frank, C. Hainzl, R. Seiringer, and J. Solovej, Microscopic derivation of Ginzburg-Landau theory, J. Am. Math. Soc. \textbf{25}, 667 (2012).
\bibitem{Larkin2005} A. Larkin and A. Varlamov, \textit{Theory of fluctuations in superconductors} (Oxford University Press, 2005).
\bibitem{Bervillier1978} C. Bervillier, J. M. Drouffe, J. Zinn-Justin, and C. Godrèche, Comparison between large-order estimates and perturbation series in a scalar field theory with Gaussian propagator, Phys. Rev. D \textbf{17}, 2144 (1978).
\bibitem{Tempere2012} J. Tempere and J. P. Devreese, Path-integral description of cooper pairing, in \textit{Superconductors: Materials, Properties and Applications} (InTech, 2012), pp. 383-414.
\bibitem{Eckern1984} U. Eckern, G. Schön, and V. Ambegaokar, Quantum dynamics of a superconducting tunnel junction, Phys. Rev. B \textbf{30}, 6419 (1984).
\bibitem{Matsubara1955} T. Matsubara, A new approach to quantum-statistical mechanics, Prog. Theor. Phys. \textbf{14}, 351 (1955).
\bibitem{Hohenberg1967} P. C. Hohenberg, Existence of long-range order in one and two dimensions, Phys. Rev. \textbf{158}, 383 (1967).
\bibitem{SanzSole2009} M. Sanz-Solé and I. Torrecilla, A fractional Poisson equation: existence, regularity and approximations of the solution, Stoch. Dyn. \textbf{9}, 519 (2009).
\bibitem{Pismen1999} L. M. Pismen, \textit{Vortices in nonlinear fields: from liquid crystals to superfluids, from non-equilibrium patterns to cosmic strings} (Oxford University Press, 1999).
\bibitem{Rubin2006} B. Rubin, Riesz potentials and integral geometry in the space of rectangular matrices, Adv. Math. \textbf{205}, 549 (2006).
\bibitem{Midei2024} G. Midei, K. Furutani, L. Salasnich, and A. Perali, Predictive power of the Berezinskii-Kosterlitz-Thouless theory based on renormalization group throughout the BCS-BEC crossover in two-dimensional superconductors, Phys. Rev. B \textbf{110}, 214502 (2024).
\bibitem{Jose2017} J. V. José, Duality, gauge symmetries, renormalization groups and the BKT Transition, Int. J. Mod. Phys. B \textbf{31}, 1730001 (2017).
\bibitem{Jose1977} J. V. José, L. P. Kadanoff, S. Kirkpatrick, and D. R. Nelson, Renormalization, vortices, and symmetry-breaking perturbations in the two-dimensional planar model, Phys. Rev. B \textbf{16}, 1217 (1977).
\bibitem{Kosterlitz2016} J. M. Kosterlitz, Kosterlitz-Thouless physics: a review of key issues, Rep. Prog. Phys. \textbf{79}, 026001 (2016).
\bibitem{Marques2024} A. M. Marques, Effects of Peierls phases in open linear chains, Phys. Rev. B \textbf{110}, 115156 (2024).
\bibitem{Prohammer1991} M. Prohammer and J. P. Carbotte, London penetration depth of d-wave superconductors, Phys. Rev. B \textbf{43}, 5370 (1991).
\bibitem{Pippard1954} A. B. Pippard, The anomalous skin effect in anisotropic metals, Proc. R. Soc. London A \textbf{224}, 273 (1954).
\bibitem{Baturina2008} T. I. Baturina, A. Bilušić, A. Y. Mironov, V. M. Vinokur, M. R. Baklanov, and C. Strunk, Quantum-critical region of the disorder-driven superconductor-insulator transition, Physica C \textbf{468}, 316 (2008).
\bibitem{Sacepe2008} B. Sacépé, C. Chapelier, T. I. Baturina, V. M. Vinokur, M. R. Baklanov, and M. Sanquer, Disorder-induced inhomogeneities of the superconducting state close to the superconductor-insulator transition, Phys. Rev. Lett. \textbf{101}, 157006 (2008).
\bibitem{Meyer2002} R. Meyer, S. E. Korshunov, C. Leemann, and P. Martinoli, Dimensional crossover and hidden incommensurability in Josephson junction arrays of periodically repeated Sierpinski gaskets, Phys. Rev. B \textbf{66}, 104503 (2002).

\end{thebibliography}
\end{document}